\documentclass[aps,prb,twocolumn,groupedaddress,showpacs]{revtex4}

\usepackage{graphicx}

\begin{document}

\title{c-Axis Transport and Resistivity Anisotropy of Lightly- to 
Moderately-Doped La$_{2-x}$Sr$_x$CuO$_4$ Single Crystals: 
Implications on the Charge Transport Mechanism}

\author{Seiki Komiya}
\email[]{komiya@criepi.denken.or.jp}
\author{Yoichi Ando}
\email[]{ando@criepi.denken.or.jp}
\author{X. F. Sun}
\author{A. N. Lavrov}

\affiliation{Central Research Institute of Electric Power Industry, 
Komae, Tokyo 201-8511, Japan}

\date{\today}

\begin{abstract}
Both the in-plane and the out-of-plane resistivities 
($\rho_{ab}$ and $\rho_c$) are measured in 
high-quality La$_{2-x}$Sr$_x$CuO$_4$ single crystals in the 
lightly- to moderately-doped region, $0.01 \le x \le 0.10$, 
and the resistivity anisotropy is determined. 
In all the samples studied, the anisotropy ratio $\rho _c/\rho _{ab}$ 
quickly increases with decreasing temperature, 
although in non-superconducting samples the strong 
localization effect causes $\rho _c/\rho _{ab}$ to decrease at 
low temperatures. 
Most notably, it is found that $\rho _c/\rho _{ab}$ at moderate 
temperatures ($100-300$ K) is almost completely independent of doping 
in the non-superconducting regime ($0.01 \le x \le 0.05$); this indicates 
that the same charge confinement mechanism that renormalizes the 
$c$-axis hopping rate is at work down to $x$=0.01. 
It is discussed that this striking $x$-independence of 
$\rho_c / \rho_{ab}$ is consistent with 
the idea that holes form a self-organized network of hole-rich regions, 
which also explains the unusually metallic in-plane transport of the holes 
in the lightly-doped region. 
Furthermore, the data for $x > 0.05$ suggest that the emergence of the 
superconductivity is related to an increase in the $c$-axis coupling. 
\end{abstract}


\maketitle

\section{Introduction}

The mechanism of the normal-state charge transport 
remains to be one of the central issues in the 
studies of the high-$T_c$ cuprates, because it is intimately tied to the 
peculiarities of the strongly-correlated electron system in which 
the high-$T_c$ superconductivity is realized.  
The contrasting behavior between the in-plane resistivity 
$\rho_{ab}$ and the $c$-axis resistivity $\rho_c$ is arguably the 
most unusual property of the charge transport in the 
cuprates (Ref. \onlinecite{Anderson}); namely, there is 
an intriguing coexistence of a metallic behavior in $\rho_{ab}(T)$ 
($d\rho_{ab}/dT > 0$) and an ``insulating" (or semiconducting) behavior 
in $\rho_{c}(T)$ ($d\rho_{c}/dT < 0$) in many 
cuprates, \cite{Nakamura,Takenaka,AndoBi2201}  
and such an unusual nature of the anisotropic charge transport appears to 
offer a key to understand the charge transport mechanism. 
In fact, already a large amount of works have been concentrated on this 
issue. \cite{Gray} 
Probably the most important feature in the $c$-axis transport is that 
the magnitude of $\rho_c(T)$ is orders of magnitude larger than that 
expected from the band-structure estimate \cite{Pickett}; 
this indicates that there is some ``charge confinement" mechanism 
\cite{Anderson} that renormalizes the $c$-axis hopping rate to a 
much smaller value.
It has been argued that such ``confinement" gives evidence for 
a non-Fermi-liquid nature of the cuprates, \cite{Anderson} 
and the persistence of the contrasting behavior between $\rho_{ab}(T)$ 
and $\rho_c(T)$ down to very low 
temperature \cite{AndoBi2201} does indeed suggest that some 
unconventional mechanism is responsible for the charge confinement, 
since any Fermi-liquid-based mechanism that provides an apparent 
confinement at finite temperatures should saturate at low enough 
temperature. \cite{Rojo,Jayannavar,Graf}  

While it is likely that some unconventional mechanism (possibly a 
non-Fermi-liquid ground state) is fundamentally responsible for the 
confinement of charges into the CuO$_2$ planes (i.e. renormalization of the 
$c$-axis hopping rate), recent studies have shown 
\cite{Ioffe,Suzuki,Krasnov} that a combination of the 
pseudogap and the $\mathbf{k}$-dependence of the $c$-axis matrix element 
\cite{Xiang} is largely responsible for the steeply insulating behavior of 
$\rho_c(T)$ observed, in particular, in Bi$_2$Sr$_2$CaCu$_2$O$_{8+\delta}$ 
(Bi-2212); namely, the angle-resolved photoemission spectroscopy (ARPES) 
measurements of Bi-2212 have demonstrated \cite{Norman} that the pseudogap 
causes destruction of the Fermi surface starting from the $(0,\pm\pi)$ 
and $(\pm\pi,0)$ points, and the $c$-axis matrix-element effect 
\cite{Ioffe,Xiang} tends to amplify the contribution of the electrons 
on these gapped portions of the Fermi surface to the $c$-axis transport.  
As a result, the $c$-axis conductivity is quickly diminished upon opening 
of the pseudogap; the steep upturn in $\rho_c(T)$ observed in 
YBa$_2$Cu$_3$O$_{7-\delta}$ (YBCO),\cite{Takenaka} 
Bi-2212,\cite{Watanabe1,Watanabe2} and 
Bi$_2$Sr$_2$CuO$_{6+\delta}$ (Bi-2201) \cite{LavrovPG} 
below the pseudogap temperature is considered to be due to this mechanism. 
On the other hand, it has been shown by the ARPES measurements that the 
Fermi surface of the La$_{2-x}$Sr$_x$CuO$_4$ (LSCO) system appears to be 
quite different from that of Bi-2212 (the stripe fluctuations cause the 
Fermi surface to be one-dimensional (1D) like \cite{Zhou,InoFS}), 
and it is not likely that the pseudogap shows a similar development 
on the Fermi surface of the LSCO; therefore it would be illuminating to 
examine the $c$-axis transport of LSCO, particularly in the 
underdoped region, and compare it with that of other cuprate systems. 

In the past, the $c$-axis transport in LSCO has already been rather 
well studied through various techniques, such as resistivity, 
\cite{Ito,Nakamura,logT,Hiroshima} magnetoresistance, 
\cite{KimuraMR,Hussey} microwaves, \cite{Shibauchi} 
and optics. \cite{Uchida,Basov} 
This is partly because large single crystals can be grown with 
the traveling-solvent floating-zone (TSFZ) technique for LSCO and thus 
it is easy to obtain samples that are long enough along the $c$-axis
to allow reliable $c$-axis transport measurements. 
It has been demonstrated \cite{Nakamura} that 
$\rho_c(T)$ in slightly underdoped 
samples ($0.10 < x < 0.16$) shows only a weak temperature dependence 
at moderate temperatures ($100 - 300$ K), particularly after correcting 
for the thermal expansion, \cite{Hiroshima} which is in contrast to the 
behaviors of $\rho_c(T)$ of Bi-2212 or Bi-2201, 
\cite{Watanabe1,Watanabe2,LavrovPG} where 
the pseudogap causes $\rho_c(T)$ to become steeply insulating. 
The magnetoresistance study by Hussey \textit{et al.} \cite{Hussey} 
has shown that the $c$-axis transport of LSCO is incoherent and 
$\rho_c$ is governed in a large part by the in-plane scattering rate.  
The low-temperature normal-state behavior of $\rho_c(T)$ 
in LSCO has been found to track the behavior of $\rho_{ab}$, namely, 
both $\rho_{ab}(T)$ and $\rho_c(T)$ for $0.08 \le x \le 0.15$ show a peculiar 
localizing behavior with the same temperature dependence of $\log (1/T)$ 
(Refs. \onlinecite{logT,Boebinger}); this seems to be in good accord with 
the magnetoresistance result of Hussey \textit{et al.}, \cite{Hussey} 
which suggests that the strong in-plane scattering can cause the carriers 
to localize in both the $ab$ and $c$ directions. 

However, all the $c$-axis transport studies that used the TSFZ-grown 
single crystals have been limited to the region with 
$x \ge 0.06$, and thus it is important to expand the region of $x$ to 
smaller values, down to the lightly-doped region, to establish a 
complete picture of the $c$-axis transport in LSCO.  
Since we have recently succeeded in growing a series of high-quality 
single crystals of LSCO from $x$ = 0.01 to 0.17 
and have measured their in-plane transport properties, \cite{mobility} 
it is natural for us to study the $c$-axis transport using those 
high-quality single crystals.
[There were studies of the transport properties of flux-grown 
La$_2$CuO$_{4+\delta}$ single crystals in the 
antiferromagnetic regime, \cite{Thio,Cheong} 
but the data are inconsistent with the present study (for $\rho_c$) and 
with Ref. \onlinecite{mobility} (for $\rho_{ab}$).] 
Studies of the $c$-axis transport in the lightly-doped LSCO would be 
particularly useful for clarifying the mechanism of the $c$-axis 
transport, in view of the unusually metallic behavior of the 
in-plane mobility which suggests that the holes are likely to form a 
self-organized network of hole-rich regions where a band-like metallic 
transport occurs. \cite{mobility} 
(Such a phase has been theoretically predicted to be realized in 
correlated electron systems and was named ``electronic liquid crystal". 
\cite{Kivelson})
Also, recent ARPES measurements \cite{InoFS} have provided information 
on the Fermi surfaces of lightly-doped LSCO, which is useful in 
interpreting the transport data. 

In this paper, we detail our technique to grow high-quality LSCO 
single crystals and present the $c$-axis transport data taken on 
the crystals with $x = 0.01 - 0.10$.  
The corresponding $\rho_{ab}(T)$ data for the same compositions 
are also presented for the analysis of the resistivity anisotropy.  
It is found that the resistivity anisotropy ratio $\rho_c / \rho_{ab}$ 
at moderate temperatures becomes almost independent of $x$ in the 
lightly-doped region; this is contrary to the common belief that the 
resistivity anisotropy should increase with decreasing doping and thus 
is a surprising result.  We discuss that this striking $x$-independence 
of $\rho_c / \rho_{ab}$ is consistent 
with the idea that holes are segregated into a network of hole-rich 
regions and that the $c$-axis transport is essentially an incoherent 
hopping between these self-organized networks, in which the strength of the 
``confinement" does not change with $x$. 

\section{Experimental Details}

\subsection{Crystal Growth and Preparation}

The series of La$_{2-x}$Sr$_x$CuO$_{4-\delta}$ single crystals 
($0.01\leq x \leq 0.10$) are grown by the TSFZ technique. 
Raw powders of La$_2$O$_3$, SrCO$_3$, and CuO (with purities of 
99.9\% or higher) are dried at 600$^\circ$C for 12 hours prior 
to weighing, and then are weighed and mixed with a composition in which 
CuO is $2-3$\% richer than the nominal one; 
the additional CuO works as a binder of the 
polycrystalline feed-rod and helps avoiding the occurrence of cracks 
in the rod, as well as avoiding penetration of the molten liquid into it, 
during the TSFZ operation. 
Also, since the vapor pressure of CuO during the growth is relatively 
high and thus the resulting crystals should have smaller Cu content 
compared to the starting material, 
the excess CuO in the feed-rod compensates the Cu loss and is necessary to 
obtain stoichiometric single crystals as the final product. 

The mixture of raw powders is well ground and calcined at 750$^\circ$C for 
12 hours, then at 920$^\circ$C for 12 hours for 4 times in alumina 
crucibles with regrinding between each calcination. 
After the last calcination and regrinding, the powders are isostatically 
pressed into a rod shape with typical dimensions of 7 mm$\phi \times$ 150 mm, 
and finally sintered at 1200$^\circ$C for 15 hours in air to form the 
feed-rod. 
Calcination and regrinding for more than 4 times before the final sintering 
make the polycrystalline materials highly homogeneous, and this 
homogeneity is important for the sintered feed-rod to become dense 
and hard, which ensures the molten zone to be well stabilized during the 
TSFZ operation. 

The solvent material, which is used to lower the temperature of the molten 
zone, is prepared to have the cation ratio of 
$\mbox{La} : \mbox{Sr} : \mbox{Cu} = (2-x) : x : 3$ for each $x$. 
We use 0.5 g of the solvent for the TSFZ growth. 
The TSFZ operation is carried out using an infrared image furnace 
(NEC Machinery SC K-15HD) with two halogen lamps and double 
ellipsoidal mirrors in the atmosphere of flowing dried air. 
To cut off reflections of the light from high angles, a quartz tube 
(which encloses the moving rods) is partly covered with aluminum foils 
to make a narrow window;
this sharpens the temperature profile and helps 
stabilizing the molten zone. \cite{Lee} 
Use of a seed crystal helps the whole rod to become a single crystal. 
The growth rate is kept constant at 1.0 mm/h (or less).  
A photograph of our TSFZ-grown crystal with $x$=0.04 is shown in 
Fig. 1(a) as an example, together with its X-ray rocking curve [Fig 1(b)]. 
The full-width-at-half-maximum (FWHM) of the (006) rocking curve 
shown in Fig. 1(b) is only 0.1$^{\circ}$, which is the smallest reported 
value for LSCO single crystals. 
(The X-ray beam is always wide enough to cover the whole specimen when 
the rocking curve is taken.) 

\begin{figure}[b]
\includegraphics[width=8.5cm]{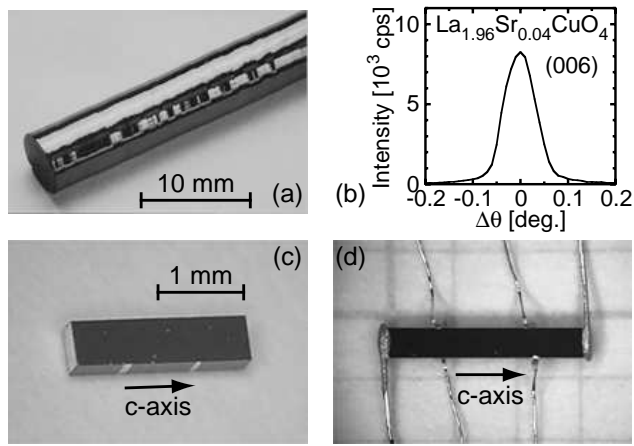}
\caption{(a) Photograph of a TSFZ-grown crystal with $x$=0.04. 
(b) X-ray (006) rocking curve of the $x$=0.04 crystal, which shows 
the FWHM of only 0.1$^{\circ}$. 
(c) Geometry of the contact pads on a $\rho_c$ sample.
(d) A $\rho_c$ sample with lead wires attached.}
\end{figure}

After the growth is finished, the whole chunk of crystal is annealed 
at 850$^\circ$C for 20 hours in air to remove thermal stresses 
formed during the FZ growth. 
The crystallographic axes are precisely determined by the X-ray 
back-reflection Laue method, which also reveals that the typical 
growth direction is $[$110$]$ 
(in the notation of high-temperature tetragonal phase). 
The samples for the present measurements are shaped into thin platelets 
with typical dimensions of 2.0 $\times$ 0.5 $\times$ 0.2 mm$^3$, 
with the wide faces exactly parallel to the $ab$ ($ac$) planes within 
an error of 1$^{\circ}$ for the in-plane (out-of-plane) measurements.
 
The samples with $x \leq 0.05$ are carefully annealed in pure Ar gas at 
400$^\circ$C for 20 hours to remove excess oxygen. 
It should be noted that in the lightly-doped region the transport 
properties are strongly affected by the excess oxygen; even a slight amount 
of excess oxygen can easily cause phase-separated oxygen-rich regions 
which show superconductivity.  (Many of the early data of 
lightly-doped LSCO, like those of Ref. \onlinecite{Takagi}, are 
actually contaminated by the superconductivity 
caused by the excess oxygen.) 
We have examined various annealing conditions and found that annealing at 
higher temperature and/or for longer time in pure Ar does not change the 
resistivity any more; therefore, we have concluded that the above annealing 
is sufficient to get rid of the excess oxygen while minimizing the 
possible decomposition. 
On the other hand, the samples with $x$ = 0.08 and 0.10 are annealed 
at 800$^\circ$C 
for 40 hours in air, followed by rapid quenching to room temperature, 
to remove oxygen defects \cite{Kanai} that cause additional electron 
scattering in moderately-doped and overdoped samples.

\subsection{Measurements}

The in-plane and out-of-plane resistivities ($\rho _{ab}$ and $\rho _c$) 
are measured using a standard ac four-probe method.  
Note that, since we can obtain samples that are long along the $c$-axis, 
we do not need to use a complicated technique, such as the Montgomery 
method, for the $\rho_c$ measurement in the case of LSCO. 
The contact pads are hand-drawn on polished sample surfaces with gold 
paint, followed by a heat treatment at 400$^\circ$C for 
30 minutes in air, which makes the gold particles to well adhere to the 
crystal surface [the high-temperature annealing to remove excess oxygen 
(or oxygen defects) are done after this process].  
For the current contacts, the whole area of two opposing side faces 
is painted with gold to ensure uniform current flow through the sample.
After the annealing, thin gold wires are attached to the contact pads 
using silver epoxy, which electrically and mechanically binds the wire 
to the sample; 
the contact resistance with this technique is less than 1 $\Omega$. 
The photographs in Figs. 1(c) and 1(d) show the geometry of the 
contacts and how the leads are attached. 

The uncertainty in the absolute values of $\rho _{ab}$ and $\rho _c$ is 
minimized by using relatively long samples (at least 2 mm long) 
and by painting narrow contact pads with a width of $50-80$ $\mu$m; 
total errors in the absolute values are less than 10\%. 
We note that, although the measurements employ a simple four-probe 
method, each sample has four voltage contacts (two on each side face) 
and the resistivity is measured using both pairs [see Fig. 1(d)].  
Sometimes the resistivity values from the two voltage pairs do not agree, 
which is an indication of some inhomogeneity (or microcrack) in the sample; 
we reject the data when such a discrepancy is observed, and all the data 
shown here are measured on samples for which the two data sets match 
within 5\%.  We also confirm that the resistivity data measured on 
different pieces of crystals from the same batch are reproducible within 
10\% as long as the samples are free from inhomogeneity.

\section{Results}

\begin{figure}[b]
\includegraphics[width=8.5cm]{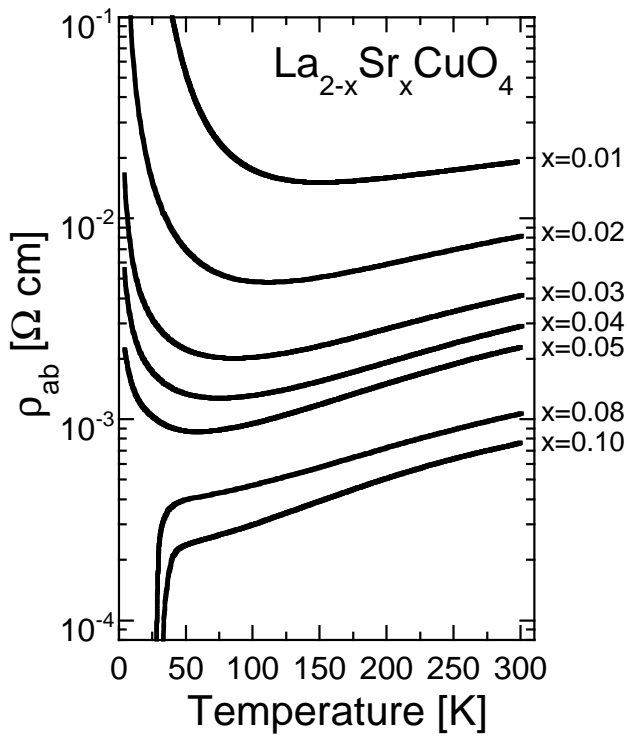}
\caption{Temperature dependences of $\rho_{ab}$ of the LSCO single 
crystals with $x$ = 0.01 -- 0.10.}
\end{figure}

Figure 2 shows the temperature dependences of $\rho _{ab}$ of our crystals 
with the vertical axis in the logarithmic scale; this figure is essentially 
a partial reproduction of the data from Ref. \onlinecite{mobility}, 
except for the $x$=0.10 data that are new. 
The absolute value of $\rho _{ab}$ is among the smallest ever reported 
\cite{KimuraMR,Sasagawa} for each $x$. 
This fact, together with the very sharp X-ray rocking curve, 
indicates that the single crystals studied here are of very high quality. 
As was emphasized in Ref. \onlinecite{mobility}, the temperature dependence 
of $\rho_{ab}$ is metallic ($d\rho _{ab}/dT > 0$) at moderate temperatures 
even in the samples where $\rho _{ab}$ exceeds 2 m$\Omega$cm, which 
corresponds to $k_Fl$ value \cite{mobility} of less than 1, and thus the 
Mott-Ioffe-Regel limit for metallic transport is strongly violated; 
we have argued \cite{mobility} that this behavior, in combination with the 
mobility that is only weakly doping-dependent, is best understood 
to result from a self-organized network of hole-rich regions that 
constitute the path for the charge transport.

Figure 3 shows the temperature dependence of $\rho _c$ for the same 
doping range. 
In the $\rho _c(T)$ profile of LSCO, a clear kink is usually observed 
at the structural phase transition temperature from the high-temperature 
tetragonal (HTT) phase to the low-temperature orthorhombic (LTO) 
phase \cite{Nakamura,Hiroshima,Hussey,Boebinger}; for all the compositions 
of the present study, however, the structural phase transition occurs at 
temperatures higher than 300 K (Ref. \onlinecite{Kastner}), except for 
the $x$=0.10 sample which show the LTO transition at 293 K. 
We note that the data for $x$=0.01 show a kink at 240 K, which corresponds 
to the N\'{e}el temperature $T_N$ and not to the LTO transition temperature. 
(The kink in the $\rho_c(T)$ curves at $T_N$ has been reported 
\cite{LavrovC1,LavrovC2} for antiferromagnetic YBCO, but not for LSCO before.)
It is useful to note that, although $\rho _c$ increases with decreasing 
temperature in all the samples, the temperature dependence of $\rho _c$ 
is rather weak at moderate temperatures (for example, $\rho _c$ of the 
$x$=0.02 sample shows only a factor of 2 increase from 300 K to 50 K), 
and a negative curvature ($d^2\rho_c/dT^2 < 0$) is observed in the 
$\rho_c(T)$ profile for $x$ = 0.02 and 0.03 at temperatures above 100 K
(which can be seen in Fig. 4).
This negative curvature is in sharp contrast to the steeply positive 
curvature of $\rho_c(T)$ of Bi-2212 (Ref. \onlinecite{Watanabe1}) or 
Bi-2201 (Ref. \onlinecite{LavrovPG}) below the pseudogap temperature.
At lower temperatures below 50 K, the $\rho_c(T)$ curves of all the samples 
show strongly diverging behavior, which is in clear correspondence with 
the localization behavior in $\rho_{ab}(T)$ at low temperatures
(except for the $x$=0.08 and 0.10 samples whose $\rho_{ab}(T)$ curves show 
an insulating behavior only when the superconductivity is suppressed 
by a high magnetic field \cite{logT}). 

\begin{figure}[b]
\includegraphics[width=8.5cm]{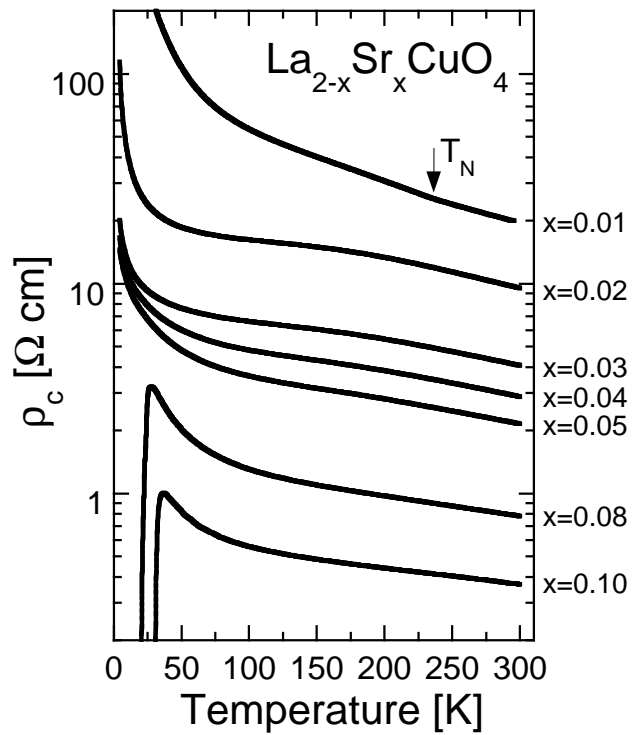}
\caption{Temperature dependences of $\rho_c$ of the LSCO single crystals 
with $x = 0.01 - 0.10$.
The arrow marks a kink at the N\'{e}el temperature for $x$=0.01.}
\end{figure}

An intriguing aspect of the $\rho _c(T)$ data is that their profile 
does not change much with $x$ (namely, the $\rho _c(T)$ data are almost 
parallel-shifted for different $x$) at moderate temperatures, as is 
the case with the $\rho _{ab}(T)$ data. 
To examine the doping dependence of the $\rho _c(T)$ in detail, 
we plot the temperature dependences of $n_{h}e\rho _c$, which corresponds 
to the inverse mobility $\mu_{c}^{-1}$ of the doped holes along the 
$c$-axis, where $e$ is the electronic charge and $n_h$ is the nominal hole 
concentration given by $2x/V$ (unit cell volume $V$ is 
3.8$\times$3.8$\times$13.2 \AA$^3$), as we did for $\rho_{ab}(T)$ in 
Ref. \onlinecite{mobility}. 
As is shown in Fig. 4, the absolute magnitude of $n_{h}e\rho _c$ at 300 K 
changes only by $\sim$40\% in the non-superconducting regime 
(from $x$ = 0.01 to 0.05), although it starts to change quickly in the 
superconducting regime (above $x$=0.05).
The weak $x$-dependence of $n_{h}e\rho _c$ at 300 K in the 
non-superconducting regime is rather similar to the behavior of the 
inverse in-plane mobility $\mu_{ab}^{-1}$ that also changes only by 
$\sim$40\% from $x$ = 0.01 to 0.05 (Ref. \onlinecite{mobility}).

\begin{figure}[b]
\includegraphics[width=8.5cm]{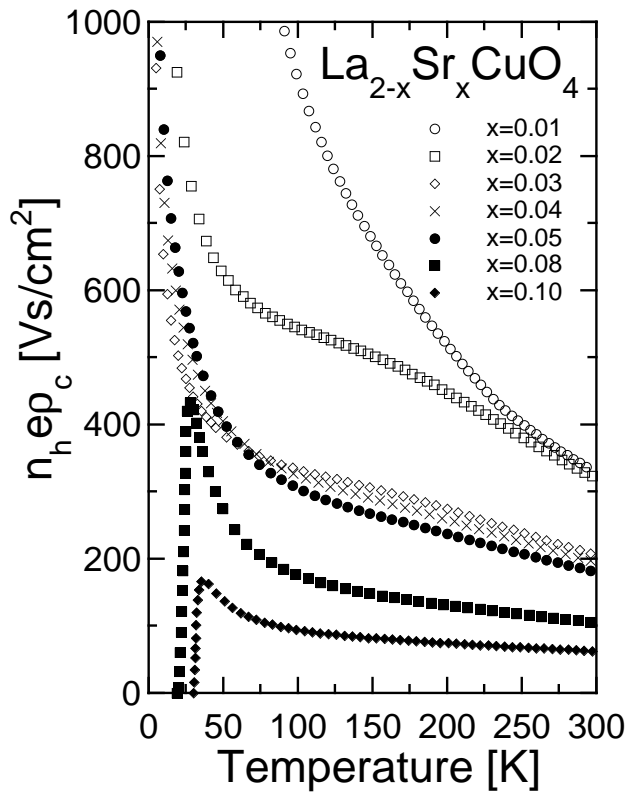}
\caption{Temperature dependences of $n_{h}e\rho _c$, which corresponds 
to the inverse mobility of the doped holes along the $c$-axis, for 
$x$ = 0.01 -- 0.10.}
\end{figure}

The most surprising feature we find in this work is that 
$\rho _c/\rho _{ab}$ at moderate temperatures is almost completely 
independent of doping in the non-superconducting regime 
($0.01 \le x \le 0.05$), as illustrated in Fig. 5. 
When $x$ exceeds 0.05 and enters into the superconducting regime, 
$\rho _c/\rho _{ab}$ starts to become smaller with increasing $x$.  
(The implication of this result is discussed in detail in the next section.)
Also, all the samples in Fig. 5 show strongly temperature-dependent 
$\rho _c/\rho _{ab}$, and in the non-superconducting samples 
($x \le 0.05$) there is a peak that moves systematically to 
higher temperature with decreasing $x$.  When one compares Fig. 5 
to Fig. 2, it becomes clear that the peak temperature of the anisotropy 
corresponds to the onset of the insulating behavior in $\rho_{ab}(T)$; 
namely, the decrease of $\rho _c/\rho _{ab}$ with decreasing temperature 
is determined by the rapid increase in $\rho_{ab}$.
The occurrence of the peak in $\rho _c/\rho _{ab}$ means that the strong 
localization of the carriers causes 
$\rho_{ab}$ to diverge more rapidly than $\rho_{c}$, and as a result 
the system becomes less two-dimensional in the strongly localized state; 
this can be rephrased that the strong localization causes the system to 
become a ``three-dimensional" insulator where the carriers cannot move to 
any direction.

\begin{figure}[b]
\includegraphics[width=8.5cm]{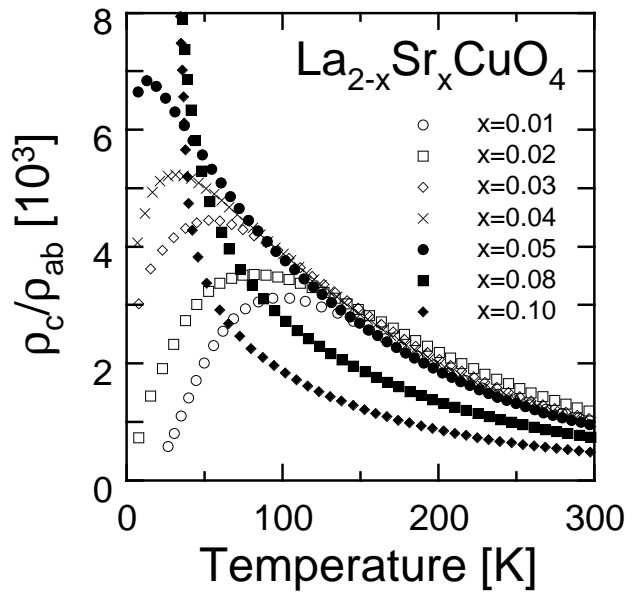}
\caption{Temperature dependences of the resistivity anisotropy ratio 
$\rho _c/\rho _{ab}$ for $x$ = 0.01 -- 0.10.}
\end{figure}

It is worth noting that it is \textit{not} the establishment of the 
long-range antiferromagnetic order but the strong localization effect 
that cause $\rho _c/\rho _{ab}$ to decrease; in this regard, our result 
disagrees with the interpretation of Kitajima \textit{et al.} 
\cite{Terasaki} for the $\rho _c/\rho _{ab}$ behavior of insulating Bi$_2$Sr$_2$ErCu$_2$O$_8$.  Our sample with $x$=0.01 clearly shows 
that $\rho _c/\rho _{ab}$ keeps increasing with decreasing temperature 
(which means that the charge confinement becomes increasingly effective) 
even below the N\'{e}el temperature until the carriers start to localize.

\section{Discussions}

As is seen in Fig. 3, the temperature dependence of $\rho_c$ 
at moderate temperatures ($100-300$ K) is weak for $x \ge 0.02$, which 
is in contrast to the $\rho_c(T)$ behaviors of Bi-2212 or Bi-2201 
where $\rho_c(T)$ shows a steep upturn 
below the pseudogap temperature [a steep upturn in $\rho_c(T)$ of LSCO 
is only observed at low temperature (below $\sim$50 K), which is 
most likely to be associated with the strong localization]. 
Since the $c$-axis transport in any cuprate should necessarily reflect 
the $c$-axis matrix element that tends to amplify the development of the 
pseudogap with a $d_{x^2-y^2}$ symmetry, the $\rho_c(T)$ behavior of 
underdoped LSCO means either that the pseudogap is already fully 
developed at 300 K, or that the pseudogap has a symmetry different 
from $d_{x^2-y^2}$.  (The photoemission experiment has observed a 
pseudogap in LSCO only in the angle-integrated mode, and the temperature 
evolution was not studied \cite{InoPG}).  
Note also that the published ARPES results to date 
have reported \cite{InoFS} that the Fermi surface in the underdoped LSCO 
is observed only near the $(\pm\pi,0)$, $(0,\pm\pi)$ points 
(1D-like Fermi surface), which is incompatible with the concept of the 
pseudogap with a $d_{x^2-y^2}$ symmetry; 
this fact might also be related to the weak temperature 
dependence of $\rho_c$ at moderate temperatures.  
(It should be noted, however, that very recent ARPES measurements 
of the lightly-doped LSCO have succeeded 
in observing a developing band at the nodal points, \cite{Fujimori} 
so the picture of the Fermi surface of LSCO is not finalized yet.)
In any case, we expect that future ARPES measurements will help 
understanding the behavior of $\rho_c(T)$ in LSCO.  

The most illuminating observation is that the resistivity anisotropy 
$\rho_c / \rho_{ab}$ near room temperature is almost independent of $x$ 
for $0.01 \le x \le 0.05$ (Fig. 5).  
In this $x$-independent regime, the magnitude of $\rho_c / \rho_{ab}$ 
is $\sim$1000 at 300 K and increases with decreasing temperature, 
which indicates that the same charge confinement mechanism is 
at work down to $x$=0.01; interestingly, this range of $x$ (0.01 -- 0.05) 
corresponds to the region where the neutron experiments have found 
diagonal spin stripes. \cite{Wakimoto,Matsuda,Matsuda2}
As was already mentioned in the previous section, the metallic in-plane 
transport that violates the Mott-Ioffe-Regel limit suggests \cite{mobility}
that the transport is occurring through a self-organized network of 
hole-rich regions \cite{Kivelson} that constitute the path for the 
charge transport. 
The striking $x$-independence of $\rho_c / \rho_{ab}$ is naturally 
understood in this picture of transport through a self-organized network,
because in such a case the transport anisotropy is determined by the 
local electronic nature of the hole-rich segment that is presumably 
unchanged with $x$ (only the average distance between the hole-rich 
paths changes with $x$).  Therefore, the behavior of $\rho_c / \rho_{ab}$ 
in the lightly-doped region is consistent with the self-organized-network 
picture and gives strong support to it.

It should be noted that the $c$-axis mobility, as well as the 
$\rho_c / \rho_{ab}$ ratio, quickly changes when $x$ enters into 
the superconducting regime.  As can be seen in Fig. 4, $n_{h}e\rho _c$ 
at 300 K changes by a factor of 3 from $x$=0.05 to 0.10, which indicates 
that the $c$-axis hopping quickly becomes easier as $x$ is increased 
above 0.05.  Correspondingly, the anisotropy ratio at moderate temperatures 
starts to show a decrease above $x$=0.05, which means that the charges are 
increasingly less confined in this doping range. 
This sharp change in the $c$-axis transport properties across $x$=0.05 
is in contrast to the smooth change of the in-plane transport properties, 
where the mobility at 300 K changes very smoothly as $\sim\sqrt{x}$ 
(Ref. \onlinecite{mobility}), neglecting the superconductor-to-insulator 
transition that occurs at low temperature.  
As we discussed in Ref. \onlinecite{mobility}, 
the fact that the in-plane mobility changes slowly and smoothly and 
is governed by a simple function of $x$ from $x$ = 0.01 to 0.17 
strongly suggests that the mechanism of the in-plane charge transport 
is essentially unchanged from the antiferromagnetic regime to 
optimum doping; since the transport in the antiferromagnetic regime 
is likely to be governed 
by a self-organized network (which can be interpreted to be a 
generalized version of the ``stripes" \cite{Kivelson}), 
the charge transport in the whole phase diagram up to optimum doping 
appears to be governed by such an 
electron self-organization, which also gives a natural picture for 
the charge confinement.  
The difference in the $x$ dependence between $\mu_c$ and 
$\mu_{ab}$ probably indicates that the $c$-axis coupling (or correlation) 
between the two-dimensional networks is more susceptible 
to the change in $x$ than the properties of the network itself.
It is intriguing to note that the emergence and increase of $T_c$ 
appear to be related to an increase in the $c$-axis coupling between the 
self-organized network. 

\section{Summary}

The $ab$- and $c$-axis resistivities are measured in the 
lightly- to moderately-doped LSCO single crystals and the 
resistivity anisotropy is analyzed to sort out the transport mechanism 
of this prototype cuprate. 
At moderate temperatures, the temperature dependence of $\rho _c$ 
in the underdoped LSCO is much weaker than that of underdoped 
Bi-2212 or Bi-2201; we discuss that this difference in the 
$\rho_c(T)$ behavior is related to the differences in both the 
Fermi-surface topology and the way how the pseudogap develops 
on the Fermi surface.
Moreover, we found that $\rho _c/\rho _{ab}$ near room temperature 
is almost completely independent of doping for $0.01 \le x \le 0.05$. 
This result supports the picture that the holes form a self-organized 
network of hole-rich path, which was originally suggested \cite{mobility} 
from the 
unusually metallic in-plane transport of the holes in the lightly-doped 
region; in this picture, it is understood that the band-like in-plane 
transport takes place 
through the self-organized network, while the $c$-axis transport is 
essentially an incoherent hopping between the networks, in which the 
strength of the confinement does not change with $x$ in the 
lightly-doped region. 
Our data for $x > 0.05$ suggest that the weakening of the $c$-axis 
charge confinement is related to the emergence of the superconductivity, 
while the in-plane transport mechanism appears to be unchanged from 
$x$=0.01 to optimum doping.

\begin{acknowledgments}
We thank A. Fujimori and S. A. Kivelson for helpful discussions, and 
T. Sasagawa and K. Kishio for valuable suggestions for the crystal growth. 
X.F.S. acknowledges support from JISTEC.
\end{acknowledgments}


\end{document}